\documentclass[letterpaper, 10 pt, conference]{ieeeconf}
\usepackage{times}

\usepackage{amsmath}
\usepackage{amsthm}
\usepackage{amssymb}
\usepackage[utf8]{inputenc}
\usepackage[T1]{fontenc}
\usepackage{graphicx}
\usepackage{enumerate}
\usepackage{xcolor}
\usepackage{comment}
\usepackage[backend=biber,style=ieee,citestyle=ieee]{biblatex}
\usepackage{multicol}
\usepackage{hyperref}
\hypersetup{
           breaklinks=true,   % splits links across lines
           colorlinks=true,   % displays links as colored text instead of blocks
           pdfusetitle=true,  % \title and \author values into pdf metadata
        }
\usepackage{tabularx}
\usepackage{makecell}
\usepackage[ruled,vlined]{algorithm2e}
\usepackage{tablefootnote}
\usepackage{multirow}
\usepackage{paralist}
\usepackage{placeins}
\usepackage[caption=false]{subfig}

\hypersetup{
    colorlinks=true,
    linkcolor=blue,
    filecolor=magenta,      
    urlcolor=cyan,
}

\newtheorem{theorem}{Theorem}
\newtheorem{proposition}[theorem]{Proposition}

\theoremstyle{definition}
\newtheorem{definition}{Definition}%[section]

\IEEEoverridecommandlockouts   

\begin{document}
% paper title
%\title{\huge Barrier States Embedded Robust Trajectory Optimization}

%\title{\huge Barrier States Embedded Iterative Dynamic Game for Safe and Robust Trajectory Optimization}

\title{\LARGE \bf Barrier States Embedded Iterative Dynamic Game for Robust and Safe Trajectory Optimization}

%\title{\huge Iterative Dynamic Game for Robust and Safe Trajectory Optimization via Embedded Barrier States}

%\title{\huge Barrier States Embedded Game Theoretic Dynamic Programming for Safe and Robust Control }
%\title{\huge Barrier States Embedded Game Theoretic Dynamic Programming for Robustly Safe Control }
%\title{\huge Barrier States Embedded Robust Dynamic Programming}
%\title{\LARGE Barrier States Embedded Safely Robust Trajectory Optimization}
%\title{\huge Robust Safety Embedded Differential Dynamic Programming through Differential Games}
%\title{\huge Game Theoretic Safety Embedded Differential Dynamic Programming}
%\title{\huge Game Theoretic Safety Embedded Differential Dynamic Programming for Robust Safe Control}
%\title{\huge Robustification of Safety Embedded Differential Dynamic Programming through Differential Games}

\author{Hassan Almubarak$^{1,4}$, Evangelos A. Theodorou$^{2}$ and Nader Sadegh$^{3}$ 
\thanks{E. A. Theodorou was supported by the National Science Foundation, CPS under Grant 1932288.} 
\thanks{$^{1}$School of Electrical and Computer Engineering} 
\thanks{$^{2}$The Daniel Guggenheim School of Aerospace Engineering}
\thanks{$^{3}$The George W. Woodruff School of Mechanical Engineering}
\thanks{Georgia Institute of Technology, Atlanta, GA, USA}
\thanks{$^{4}$ Department of Control and Instrumentation Engineering}
\thanks{King Fahd University of Petroleum \& Minerals, Dhahran, Saudi Arabia}
\thanks{{\tt\footnotesize halmubarak, evangelos.theodorou,sadegh@gatech.edu}}      
}
\maketitle
\thispagestyle{empty}
\pagestyle{empty}

\begin{abstract}
Considering uncertainties and disturbances is an important, yet challenging, step in successful decision making. The problem becomes more challenging in safety-constrained environments. In this paper, we propose a robust and safe trajectory optimization algorithm through solving a constrained min-max optimal control problem. The proposed method leverages a game theoretic differential dynamic programming approach with barrier states to handle parametric and non-parametric uncertainties in safety-critical control systems. Barrier states are embedded into the differential game's dynamics and cost to portray the constrained environment in a higher dimensional state space and certify the safety of the optimized trajectory. Moreover, to find a convergent optimal solution, we propose to perform line-search in a Stackleberg (leader-follower) game fashion instead of picking a constant learning rate. The proposed algorithm is evaluated on a velocity-constrained inverted pendulum model in a moderate and high parametric uncertainties to show its efficacy in such a comprehensible system. The algorithm is subsequently implemented on a quadrotor in a windy environment in which sinusoidal wind turbulences applied in all directions. 

\end{abstract}
\vspace{-1mm}
\section{Introduction}
Optimal control has been a central element in designing practical and successful decision polices including those relying upon reinforcement learning approaches for complex dynamical systems. Regardless of the employed methodology, sound decision polices must take uncertainties and disturbances into consideration. This is particularly true for safety-critical systems and especially when learning, models or policies, is involved.

Min-max optimal control, which can be considered as an ${\rm{H}}_{\infty}$ optimal control technique, is a viable robust control methodology which has been proven theoretically and practically to handle various systems' uncertainties such as model mismatch, signals noise and disturbances \cite{van19922L2andLinfControl,zhou1996robust}. The Min-max approach has a game-theoretic interpretation in which two non-cooperative players have opposing objectives \cite{Ball1989h_infandgames}. Specifically, it is a game in which a player is to minimize some payoff function while the other is to maximize.

Differential Dynamic Programming (DDP), a second order trajectory optimization technique, is an effective technique to optimize high dimensional systems' trajectories and to improve reinforcement learning outputs. With the goal of collecting meaningful training data for reinforcement learning, Morimoto et al. \cite{morimoto2003minimax} proposed the first discrete min-max DDP to provide robust control polices. Independently, and around the same time,  Ogunmolu et al. \cite{Ogunmolu2018minmaxiDG} and Sun et al. \cite{sun2018minmax} proposed a correction in the value function's recursions in the min-max DDP of Morimoto et al. \cite{morimoto2003minimax} with applications to robust nonlinear controls in \cite{Ogunmolu2018minmaxiDG} and extensions to continuous time min-max DDP in \cite{sun2018minmax}. Nonetheless, there has been no attempt to consider min-max DDP for safety-critical or constrained systems, which is a necessary extension. 

Safety of dynamical systems can be verified through forward invariance of the desired set of allowed states \cite{blanchini1999set}. Barrier-like based methods such as barrier certificates \cite{prajna2003barrier,prajna2004safety}, control barrier functions (CBFs) \cite{wieland2007constructive,ames2014control,romdlony2014uniting} and barrier states \cite{Almubarak2021SafetyEC,almubarak2021safeddp}, use barrier functions, well known in optimization, to show or enforce invariance. Utilizing barrier functions, the barrier states (BaS) method, firstly introduced in \cite{Almubarak2021SafetyEC} for safe stabilization of continuous time systems, enforces forward invariance of the safe set through augmenting the \textit{state} of the barrier function into the model of the control dynamical system converting the safety-critical control problem into a control design problem that seeks a stabilizing control law for the augmented model guaranteeing boundedness of the barrier states. This concept was adopted by the authors in the context of trajectory optimization for discrete time systems developing discrete barrier states in \cite{almubarak2021safeddp}, which was shown to consistently outperforms the penalty methods and CBFs safety filters in providing safe optimal trajectories.

\begin{figure} [tb]
    \vspace{-7mm}
    \centering
     \hspace*{-.40cm}\subfloat{\includegraphics[trim=70 5 20 20, clip, width=.7\linewidth]{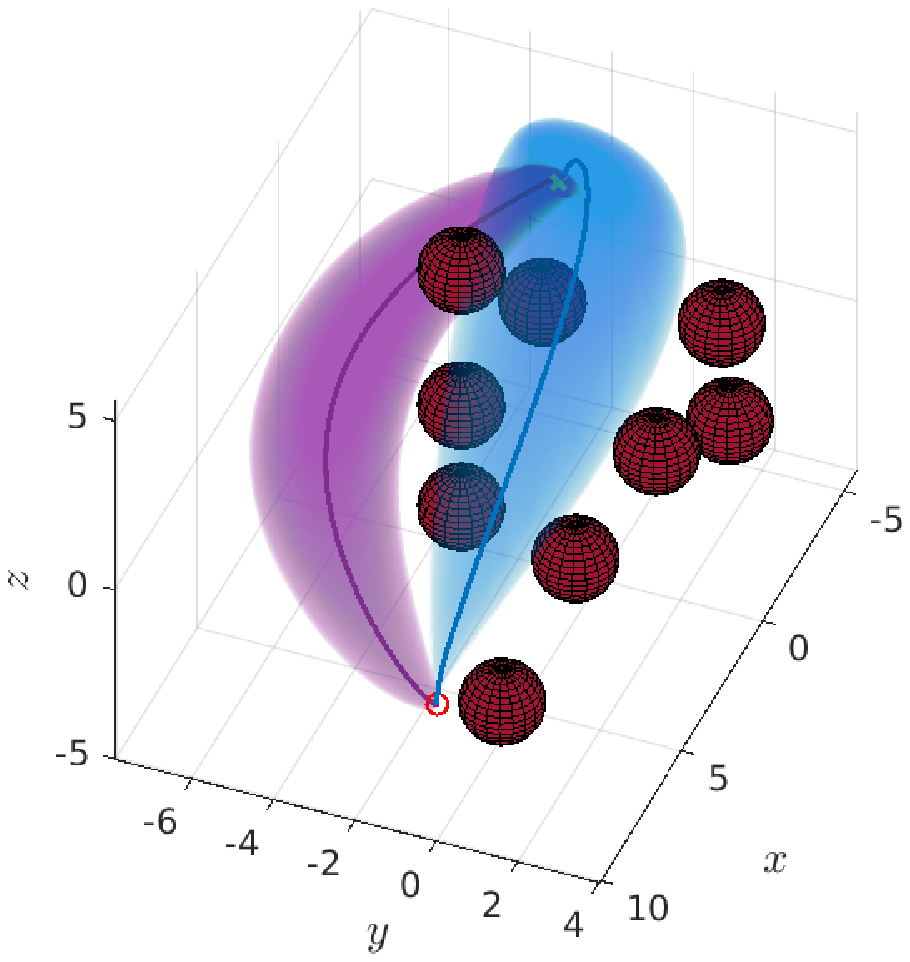}}
     \hspace*{-11mm} 
     \subfloat{\includegraphics[trim=90 5 20 10, clip, width=.6\linewidth]{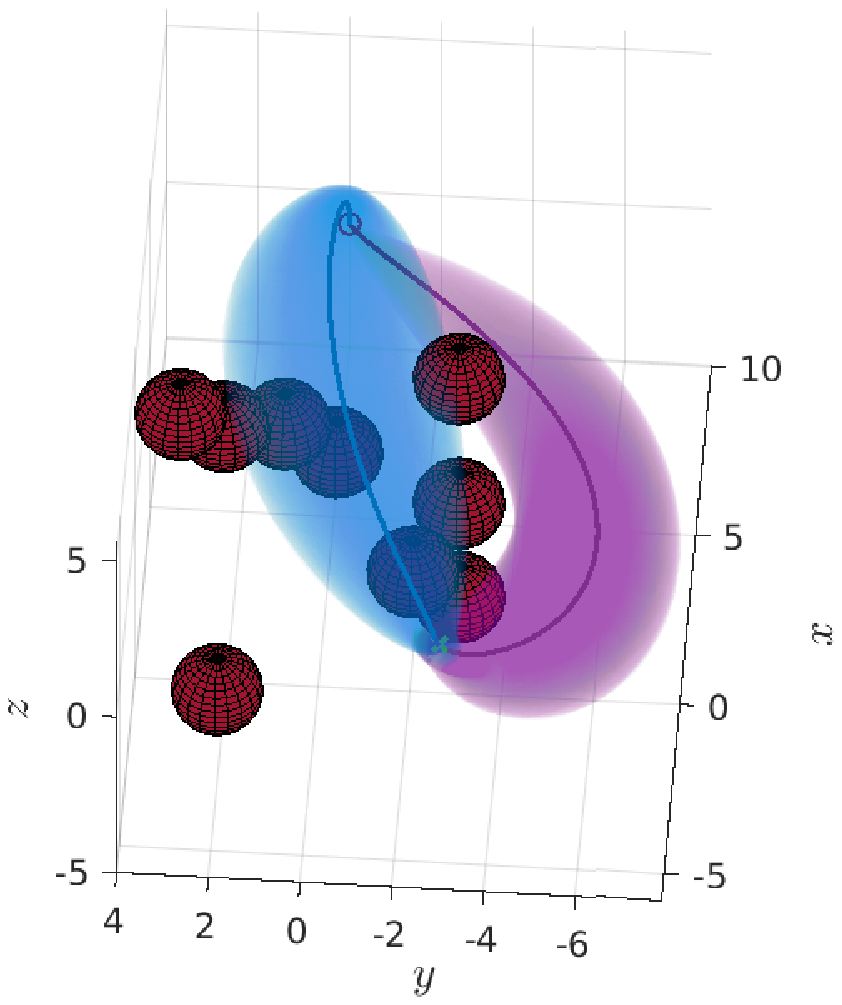}} \\
     %\hspace*{-.5cm}\subfloat{\includegraphics[trim=60 5 20 20, clip, width=.7\linewidth]{Figures/quad_two_robustness_sols_99_10.eps}}
     %\hspace*{-1.5cm}
     %\subfloat{\includegraphics[trim=60 5 20 20, clip, width=.7\linewidth]{Figures/quad_two_robustness_sols_99_11.eps}} 
     \caption{Barrier states embedded MinMax DDP successfully and safely drives the quadrotor to reach the target (green x) starting from the red circle under a Gaussian sinusoidal wind turbulence applied in all directions with a standard deviation of $15$. Shown here are two robustness results, $R_{\rm{v}}=\frac{1}{100} I$ (blue) and $R_{\rm{v}}= \frac{1}{150} I$ (purple), in which the solid trajectories represent the undisturbed trajectories and the shaded regions represent confidence regions of $95\%$.} %It can be seen that both regions are clear from obstacles but the control policy generated by the min player with less penalization of the max player takes a larger side turn away from the obstacles to be able to handle larger disturbances.}
     \label{Fig:Quadrotor with two different Rv under wind sin distrubances}
     \vspace{-5.7mm}
\end{figure}

\subsection{Contributions and Organization of the paper}
In this paper, we start by introducing some preliminaries about min-max optimal control and the associated Hamilton-Jacobi-Bellman-Isaacs partial differential equation and discrete barrier states used to enforce safety in trajectory optimization in \autoref{Sec:Preliminaries}. In \autoref{Sec:min-max safe DDP}, we use barrier states with min-max DDP to develop a safety embedded min-max DDP that provides robust and safe nonlinear control for safety-critical systems. Leveraging min-max optimal control assumptions, the developed algorithm utilizes a Stackleberg game strategy which helps finding the optimal strategies for each player, increasing the robustness of the min player that produces the feedback control policy of interest. We implement the proposed algorithm in \autoref{Section: Simulation Examples} first on a parametrically uncertain model of an inverted pendulum in which we consider two different cases of uncertainty levels. To show the efficacy of the algorithm on non-parametric disturbances, we implement it on a quadrotor flying in an obstacle course in a windy environment (\autoref{Fig:Quadrotor with two different Rv under wind sin distrubances}). Similar to the pendulum, we consider two different cases of disturbance levels. The developed algorithm is shown to consistently outperform standard DBaS-DDP \cite{almubarak2021safeddp} in safely completing the task with a much less variance in the states at the expense of having a larger root-mean-square deviation (RMSD) from the target. Finally, we provide concluding remarks and future directions in \autoref{Section: Conclusion}. %Furthermore, it is shown that a more robust performance is achieved when using the proposed leader-follower line-search against using the standard min-max DDP \cite{morimoto2003minimax,Ogunmolu2018minmaxiDG,sun2018minmax}. 

\vspace{-1mm}
\section{Preliminaries} \label{Sec:Preliminaries}
\subsection{Differential Games and Min-Max Optimal Control}
Consider the differential game problem  
\begin{equation} \label{game cost}
    J({\rm{x}}, {\rm{U}}, {\rm{V}}) =  \min_{\rm{u}} \max_{\rm{v}} \sum_{k=1}^{N-1} \mathcal{L} ({\rm{x}}_k, {\rm{u}}_k,{\rm{v}}_k) + \phi({\rm{x}}_N)
\end{equation}
subject to
\begin{equation} \label{game discrete control system}
    {\rm{x}}_{k+1}=f({\rm{x}}_k,{\rm{u}}_k,{\rm{v}}_k)
\end{equation}
where $k \in \mathbb{Z}^+_0$ is the time step, ${\rm{x}}_k \in \mathcal{X} \subset \mathbb{R}^n$ is the state of the system at time step $k$, ${\rm{u}}_k \in \mathcal{U} \subset \mathbb{R}^{m_{\rm{u}}}$ is the minimizing player, ${\rm{v}}_k \in \mathcal{V} \subset \mathbb{R}^{m_{\rm{v}}}$ is the maximizing player, ${\rm{U}}$ and ${\rm{V}}$ are the minimzing and maximizing control sequences, $\mathcal{L}:\mathbb{R}^n \times \mathbb{R}^{m_{\rm{u}}} \times \mathbb{R}^{m_{\rm{v}}} \rightarrow \mathbb{R}^{+}$ is a running cost, $\phi:\mathbb{R}^n \rightarrow \mathbb{R}^{+}$ is a terminal cost and $f:\mathcal{X}\rightarrow \mathcal{X}$ is the model of some dynamical system. In control theory, this problem is classically looked at as an optimal control problem with some disturbance or uncertainty in the dynamics in which it is desired to find an optimal control policy that is robust against disturbances and uncertainties. This problem is also known as a min-max optimal control problem. 

In optimal control, it is well known that under certain conditions, the optimal solution, known as the value function $V$, satisfies the Hamilton-Jacobi-Bellman (HJB) partial differential equation (PDE). In min-max optimal control, i.e. for the game in \eqref{game cost}-\eqref{game discrete control system}, the corresponding PDE is known as the Hamilton-Jacobi-Bellman-Isaacs (HJBI) PDE \cite{bacsar1998dynamic} which is given by 
\begin{equation}
    V({\rm{x}}_k) = \min_{{\rm{u}}_k} \max_{{\rm{v}}_k} \big\{\mathcal{L} ({\rm{x}}_k, {\rm{u}}_k,{\rm{v}}_k)+ V({\rm{x}}_{k+1}) \big\}
\end{equation}

\subsection{Barrier States}
%Safety of dynamical systems can be verified through forward invariance of the desired set of allowed states \cite{blanchini1999set}. Barrier-like based methods such as barrier certificates \cite{prajna2003barrier,prajna2004safety}, control barrier functions \cite{wieland2007constructive,ames2014control,romdlony2014uniting} and barrier states \cite{Almubarak2021SafetyEC,almubarak2021safeddp}, use barrier functions, well known in optimization, to show or enforce invariance. 
In essence, a barrier function can be defined as a continuous real valued function on a non-empty open set whose value approaches infinity as its independent variable goes close to the boundaries of the set's complement. Namely, consider a superlevel set $\mathcal{S} \subset \mathbb{R}^n$ defined by a smooth real valued function $h: \mathcal{X} \subset \mathbb{R}^n \rightarrow \mathbb{R}$ such that the set, its interior and its boundary are defined respectively as $\mathcal{S} = \{{\rm{x}}  \in \mathcal{X}: h({\rm{x}}) \geq 0\},\ \mathcal{S}^{\circ} = \{{\rm{x}}  \in \mathcal{X}: h({\rm{x}}) > 0\}$ and $\partial \mathcal{S} = \{{\rm{x}}  \in \mathcal{X}: h({\rm{x}}) = 0\}$.
%\begin{align}  \label{safe set S} 
%\begin{split}
%    \mathcal{S} &= \{{\rm{x}}  \in \mathcal{X}: h({\rm{x}}) \geq 0\}\\
%    \mathcal{S}^{\circ} &= \{{\rm{x}}  \in \mathcal{X}: h({\rm{x}}) > 0\}\\
%    \partial \mathcal{S} &= \{{\rm{x}}  \in \mathcal{X}: h({\rm{x}}) = 0\}
%\end{split}
%\end{align}
We can define a barrier function $B: \mathbb{R}^n \rightarrow \mathbb{R}$ to be a smooth function such that for some ${\rm{x}} \in \mathcal{S}^{\circ}$, if ${\rm{x}} \rightarrow  \partial \mathcal{S}$ and $h \rightarrow 0$, then $B(h) \rightarrow \infty$. $B$ is commonly selected to be a logarithmic barrier or an inverse barrier. 

Consider the discrete-time nonlinear safety-critical control system
\begin{equation} \label{discrete control system}
    {\rm{x}}_{k+1}=f({\rm{x}}_k,{\rm{u}}_k)
\end{equation}
whose states are desired to stay in the interior of the superlevel set $\mathcal{S}$  %defined in \eqref{safe set S}
under the feedback controller ${\rm{u}}_k=K({\rm{x}}_k)$. That is, we seek to render $\mathcal{S}^{\circ}$ controlled invariant with respect to the closed loop system $f({\rm{x}}_k,K({\rm{x}}_k))$. 
\begin{definition}  \label{invariant def}
The set $\mathcal{S}^{\circ} \subset \mathbb{R}^n$ is controlled invariant, also referred to as \textit{safe}, with respect to the control dynamical system \eqref{discrete control system} if $\forall {\rm{x}}_0 \in \mathcal{S}^{\circ}$, given the feedback policy ${\rm{u}}_k=K({\rm{x}}_k)$, ${\rm{x}}_k \in \mathcal{S}^{\circ} \ \forall k\in \mathbb{Z}^+$. Equivalently, the safety condition
\begin{equation} \label{safety condition}
    h({\rm{x}}_k) > 0 \ \forall k \geq 0 ; \ {\rm{x}}(0) \in \mathcal{S}^{\circ} 
\end{equation}
is satisfied. 
\end{definition}

The barrier states (BaS) method \cite{Almubarak2021SafetyEC} enforces safety by embedding the state of the barrier function into the model of the safety-critical system expressing hard constraints as states of the system to be driven and stabilized with the original states. In other words, the safety constraints are transformed into performance objectives in a higher dimensional state space. It is worth noting that in optimal control settings, this also elevates the dimension of the sought-after value function. 

%in a higher dimensional statespace
%Utilizing barrier functions, the barrier states (BaS) method, introduced in \cite{Almubarak2021SafetyEC} for safe stabilization of continuous time systems, enforces forward invariance of the safe set through augmenting the \textit{state} of the barrier function into the model of the control dynamical system converting the safety-critical control problem into an unconstrained control design problem that seek a control law that stabilizes the augmented model guaranteeing boundedness of the barrier state which implies that the safety-critical system is safe. This concept was adopted by the authors in the context of trajectory optimization for discrete time systems, developing discrete barrier states \cite{almubarak2021safeddp}.

As $B$ can be picked to be any valid barrier function, we define the barrier function of ${\rm{x}}$ to be $\beta({\rm{x}}_k):=B \circ h({\rm{x}}_k)$. To achieve forward invariance of the set $\mathcal{S}^{\circ}$, define the barrier state $w_k:= \beta_k - \beta^d$, where $\beta_k = \beta({\rm{x}}_k)$ and $\beta^d = B \circ h({\rm{x}}^d)$ for the target state ${\rm{x}}^d$. As noted in \cite{almubarak2021safeddp}, shifting $w$ by $\beta^d$ is not required specifically for our development of safe trajectory optimization. % but it ensures that the minimum of the barrier lies at the target point which can come in handy in some applications.
From \autoref{invariant def} and the definition of barrier functions, the following proposition states a necessary and sufficient condition to enforce safety \cite{almubarak2021safeddp}. 

\begin{proposition} \label{prop:safety}
The safe set $\mathcal{S}^{\circ}$ is controlled invariant through the feedback control law ${\rm{u}}_k=K({\rm{x}}_k)$ if and only if $w_0<\infty \Rightarrow w_k<\infty \ \forall k \in \mathbb{Z}^+$.
\end{proposition}
\noindent In other words, as long as the barrier state is rendered bounded, the system is safe. Hence, the discrete barrier state (DBaS) of the safety condition \eqref{safety condition} for the system \eqref{discrete control system} can be given as
\begin{equation} \begin{split}
    %\Delta \beta(x_k) & = \beta(x_{t+1}) - \beta(x_k) = B(h(f(x_k,u_k)) - \beta(x_k) \\
     w_{t+1} & =  B \circ h(f({\rm{x}}_t,{\rm{u}}_t))  - \beta^{d} \\
\end{split} \end{equation}

\section{Robust and Safe Trajectory Optimization} \label{Sec:min-max safe DDP}
In this paper, we wish to design a safe and robust feedback control policy ${\rm{u}}$ with the existence of an invasive player ${\rm{v}}$. To solve this problem, we consider the min-max optimal control problem \eqref{game cost}-\eqref{game discrete control system} in a constrained environment, i.e. the dynamics of the system is subject to some safety constraints. Specifically, the min-max optimal control problem is subject to $h({\rm{x_k}})>0 \ \forall k \in [0, N]$. To address the safety constraint, we use discrete barrier states to portray the safety condition in the optimization problem which is then solved through differential dynamics programming (DDP). In our development, we do not impose other than standard regularity conditions on the optimization problem and the safety constraints. Namely, the dynamics of the system $f$ and the function $h$ defining the safe set are continuously differentiable, the running and terminal costs are at least twice continuously differentiable and the players polices are continuous. 

\subsection{Game Theoretic Discrete Time Barrier States}
To address the safety constrained min-max optimal control problem, we reformulate the DBaS according to the differential dynamics in \eqref{game discrete control system}. Consequently, we derive the game theoretic discrete barrier state equation (GT-DBaS) as
\begin{equation} \begin{split} \label{dbas dynamics}
    %\Delta \beta(x_k) & = \beta(x_{k+1}) - \beta(x_k) = B(h(f(x_k,u_k)) - \beta(x_k) \\
     w_{k+1} & =  B(h(f({\rm{x}}_k,{\rm{u}}_k,{\rm{v}}_k))  - \beta^{d} \\
\end{split} \end{equation}
For $q$ constraints, depending on the problem settings, one could define a single barrier state or multiple barrier states for multiple constraints. Let ${\rm{w}} \in \mathbb{R}^q$ be a vector of $q$ barrier states. Augmenting this state vector to the differential games dynamical system's states, we get $\hat{{\rm{x}}}= \begin{bmatrix} {\rm{x}} \\ {\rm{w}} \end{bmatrix}$. Therefore, the safety embedded differential game becomes
\begin{equation} \label{safety embedded game cost}
    J(\hat{{\rm{x}}},{\rm{U}},{\rm{V}}) =  \min_{\rm{u}} \max_{\rm{v}} \sum_{k=1}^{N-1} \mathcal{L} (\hat{{\rm{x}}}_k, {\rm{u}}_k,{\rm{v}}_k) + \phi(\hat{{\rm{x}}}_N)
\end{equation}
subject to the dynamics
\begin{equation} \label{safety embedded game discrete control system}
    \hat{{\rm{x}}}_{k+1}=\hat{f}(\hat{{\rm{x}}}_k,{\rm{u}}_k,{\rm{v}}_k)
\end{equation}
where $\hat{f} = \begin{bmatrix} f({\rm{x}}_k,{\rm{u}}_k,{\rm{v}}_k) \\ B(h(f({\rm{x}}_k,{\rm{u}}_k,{\rm{v}}_k))  - \beta^{d} \end{bmatrix}$. It is worth noting that the GT-DBaS can also be driven by the adversarial control $\rm{v}$ and is part of the cost function posing more difficulties to the minimizing control $\rm{u}$ in achieving performance and safety objectives in which the latter is achieved by ensuring boundedness of the GT-DBaS. As a result, unlike the unconstrained case, picking $R_{\rm{v}}$ just slightly bigger than $R_{\rm{u}}$ may not result in a convergent solution as the hostile player also has access to the safety critical state's dynamics and cost. Hence, one needs to tune $R_{\rm{v}}$ with a relatively high penalty to get a convergent solution as we will show in the application examples in \autoref{Section: Simulation Examples}.

\subsection{Safety Embedded Min-Max DDP} \label{Section: Safety Embedded Differential Dynamic Programming (DDP)}
Aiming to develop robust control policies for high-dimensional discrete-time systems, Morimoto et al. \cite{morimoto2003minimax} used discrete DDP in a min-max framework. The developed optimal policy generated by the min-max DDP was implemented on a simulated biped robot and shown to outweigh a hand-tuned PD controller which failed to handle unknown disturbances as did the standard DDP. Sun et al. \cite{sun2018minmax} extended the min-max DDP to continuous time systems and included more terms missed in the previous attempt in the DDP algorithm for the discrete case. The developed algorithm, named game-theoretic DDP (GT-DDP), was experimented on a quadrotor with a sling load which can lead to major errors in the model due to the pendulous oscillation during flight.

In this work, we are interested in robustifying the trajectory optimization problem in safety-critical environments. In particular, we are to develop a safe and robust optimal control policy through the use of DBaS-DDP in \cite{almubarak2021safeddp} in the framework of differential games, i.e. using min-max DDP. %Therefore, we follow the same min-max DDP derivations but for the safety embedded differential game \eqref{safety embedded game cost}-\eqref{safety embedded game discrete control system}.

Consider a nominal trajectory of the safety embedded states and the players policies $(\bar{\hat{\rm{x}}}, \bar{\rm{u}}, \bar{\rm{v}})$, and the safety embedded HJBI equation
\begin{equation} \label{safety embedded HJBI}
    V(\hat{\rm{x}}_t) = \min_{{\rm{u}}_t} \max_{{\rm{v}}_t} \big\{\mathcal{L} (\hat{\rm{x}}_t, {\rm{u}}_t,{\rm{v}}_t)+ V(\hat{\rm{x}}_{t+1}) \big\}
\end{equation}
with a boundary condition $V(\hat{\rm{x}}_N) = \phi(\hat{\rm{x}}_N) $.
%\begin{equation}
%    V(\hat{\rm{x}}_N) = \phi(\hat{\rm{x}}_N) 
%\end{equation}
The algorithm consists of iterative backward passes along the value function and its derivatives along the system's states resulting from expanding the HJBI equation \eqref{safety embedded HJBI} around the state-inputs nominal trajectory and forward passes along the  safety embedded dynamics \eqref{safety embedded game discrete control system}. In consideration of that, given a nominal trajectory, we compute the local second order model of the variation function $\mathcal{H}$ resulted from expanding the HJBI equation as \begin{equation}
    \begin{split}
        & \mathcal{H}_{\hat{\rm{x}}} = \mathcal{L}_{\hat{\rm{x}}} + V_{\hat{\rm{x}}} \hat{f}_{\hat{\rm{x}}} , \ \mathcal{H}_{\hat{\rm{x}}\hat{\rm{x}}} = \mathcal{L}_{\hat{\rm{x}}\hat{\rm{x}}} + \hat{f}^{\rm{T}}_{\hat{\rm{x}}} V_{\hat{\rm{x}}\hat{\rm{x}}} \hat{f}_{\hat{\rm{x}}} + V_{\hat{\rm{x}}} f_{\hat{\rm{x}}\hat{\rm{x}}}\\
        & \mathcal{H}_{{\rm{u}}} = \mathcal{L}_{{\rm{u}}} + V_{\hat{\rm{x}}} \hat{f}_{{\rm{u}}}, \ \mathcal{H}_{{\rm{uu}}} = \mathcal{L}_{{\rm{uu}}} + \hat{f}^{\rm{T}}_{{\rm{u}}} V_{\hat{\rm{x}}\hat{\rm{x}}} \hat{f}_{{\rm{u}}} + V_{\hat{\rm{x}}} f_{{\rm{u}}{\rm{u}}} \\
        & \mathcal{H}_{{\rm{v}}} = \mathcal{L}_{{\rm{v}}} + V_{\hat{\rm{x}}} \hat{f}_{{\rm{v}}}, \ \mathcal{H}_{{\rm{vv}}} = \mathcal{L}_{{\rm{vv}}} + \hat{f}^{\rm{T}}_{{\rm{v}}} V_{\hat{\rm{x}}\hat{\rm{x}}} \hat{f}_{{\rm{v}}} + V_{\hat{\rm{x}}} f_{{\rm{v}}{\rm{v}}} \\
        & \mathcal{H}_{\hat{\rm{x}}{\rm{u}}} = \mathcal{L}_{\hat{\rm{x}}{\rm{u}}} + \hat{f}^{\rm{T}}_{\hat{\rm{x}}} V_{\hat{\rm{x}}\hat{\rm{x}}} \hat{f}_{{\rm{u}}} + V_{\hat{\rm{x}}} f_{\hat{\rm{x}}{\rm{u}}}, \ \mathcal{H}_{{\rm{u}} \hat{\rm{x}}} = \mathcal{H}_{\hat{\rm{x}}{\rm{u}}}^{\rm{T}} \\
         & \mathcal{H}_{\hat{\rm{x}}{\rm{v}}} = \mathcal{L}_{\hat{\rm{x}}{\rm{v}}} + \hat{f}^{\rm{T}}_{\hat{\rm{x}}} V_{\hat{\rm{x}}\hat{\rm{x}}} \hat{f}_{{\rm{v}}} + V_{\hat{\rm{x}}} f_{\hat{\rm{x}}{\rm{v}}}, \ \mathcal{H}_{{\rm{v}}\hat{\rm{x}}} = \mathcal{H}_{\hat{\rm{x}}{\rm{v}}}^{\rm{T}} \\
         & \mathcal{H}_{{\rm{uv}}} = \mathcal{L}_{{\rm{uv}}} + \hat{f}^{\rm{T}}_{{\rm{u}}} V_{\hat{\rm{x}}\hat{\rm{x}}} \hat{f}_{{\rm{v}}} + V_{\hat{\rm{x}}} f_{{\rm{u}}{\rm{v}}}, \ \mathcal{H}_{{\rm{vu}}} = \mathcal{H}_{{\rm{uv}}}^{\rm{T}} 
        \end{split} \label{H matrices}
\end{equation}
Following the derivations in \cite{sun2018minmax} for the safety embedded min-max problem, the optimal polices are then computed as
\begin{equation} \begin{split}
     \delta {\rm{u}}_k^* = \mathbf{k}_{{\rm{u}}_k} + \mathbf{K}_{{\rm{u}}_k} \delta \hat{\rm{x}}_k , \
     \delta {\rm{v}}_k^* = \mathbf{k}_{{\rm{v}}_k} + \mathbf{K}_{{\rm{v}}_k} \delta \hat{\rm{x}}_k
\end{split} \end{equation}
where at time instant $k$,
\begin{equation*} \begin{split}
    & \mathbf{k}_{\rm{u}} = -\widetilde{\mathcal{H}}_{{\rm{uu}}}^{-1} \big( \mathcal{H}_{{\rm{u}}} - \mathcal{H}_{{\rm{uv}}} \mathcal{H}_{{\rm{vv}}}^{-1} \mathcal{H}_{{\rm{v}}}\big) \\
    & \mathbf{K}_{\rm{u}} = -\widetilde{\mathcal{H}}_{{\rm{uu}}}^{-1} \big( \mathcal{H}_{ux} - \mathcal{H}_{{\rm{uv}}} \mathcal{H}_{{\rm{vv}}}^{-1} \mathcal{H}_{vx}\big) \\
    & \mathbf{k}_{\rm{v}} = -\widetilde{\mathcal{H}}_{{\rm{vv}}}^{-1} \big( \mathcal{H}_{{\rm{v}}} - \mathcal{H}_{{\rm{vu}}} \mathcal{H}_{{\rm{uu}}}^{-1} \mathcal{H}_{{\rm{u}}}\big) \\
    & \mathbf{K}_{\rm{v}} = -\widetilde{\mathcal{H}}_{{\rm{vv}}}^{-1} \big( \mathcal{H}_{vx} - \mathcal{H}_{{\rm{vu}}} \mathcal{H}_{{\rm{uu}}}^{-1} \mathcal{H}_{ux}\big)
\end{split} \end{equation*}
given
\begin{equation*}
    \begin{split}
         \widetilde{\mathcal{H}}_{{\rm{uu}}} = \mathcal{H}_{{\rm{uu}}} - \mathcal{H}_{{\rm{uv}}} \mathcal{H}_{{\rm{vv}}}^{-1} \mathcal{H}_{{\rm{vu}}} , \
         \widetilde{\mathcal{H}}_{{\rm{vv}}} = \mathcal{H}_{{\rm{vv}}} - \mathcal{H}_{{\rm{vu}}} \mathcal{H}_{{\rm{uu}}}^{-1} \mathcal{H}_{{\rm{uv}}}
    \end{split}
\end{equation*}
Accordingly, the value function's equations used in the backward propagation are
\begin{align*} \begin{split}
\label{Backward Propagation equations of V}
& V_k = V_{k+1} + \mathbf{k}_{\rm{u}}^{\rm{T}} \mathcal{H}_{\rm{u}} + \mathbf{k}_{\rm{v}}^{\rm{T}} \mathcal{H}_{\rm{v}} + \mathbf{k}_{\rm{u}}^{\rm{T}} \mathcal{H}_{{\rm{uv}}} \mathbf{k}_{\rm{v}} \\ 
& \ \ \ \ \ \ \ \ \ \ \ \ \ + \frac{1}{2} \big( \mathbf{k}_{\rm{u}}^{\rm{T}} \mathcal{H}_{{\rm{uu}}} \mathbf{k}_{\rm{u}} + \mathbf{k}_{\rm{v}}^{\rm{T}} \mathcal{H}_{{\rm{vv}}} \mathbf{k}_{\rm{v}}  \big) \\
& V_{\hat{\rm{x}}_k}= \mathcal{H}_{\hat{\rm{x}}_k} + \mathbf{K}^{\rm{T}}_{\rm{u}} \mathcal{H}_{\rm{u}} + \mathbf{K}^{\rm{T}}_{\rm{v}} \mathcal{H}_{\rm{v}} + \mathcal{H}_{\hat{\rm{x}} {\rm{u}}} \mathbf{k}_{\rm{u}} +  \mathcal{H}_{\hat{\rm{x}} {\rm{v}}} \mathbf{k}_{\rm{v}} \\
& \ \ \ \ \ \ \ + \mathbf{K}^{\rm{T}}_{\rm{u}} \mathcal{H}_{{\rm{uu}}} \mathbf{k}_{\rm{u}}  + \mathbf{K}^{\rm{T}}_{\rm{v}} \mathcal{H}_{{\rm{vv}}} \mathbf{k}_{\rm{v}} + \mathbf{K}^{\rm{T}}_{\rm{u}} \mathcal{H}_{{\rm{uv}}} \mathbf{k}_{\rm{v}} + \mathbf{K}^{\rm{T}}_{\rm{v}} \mathcal{H}_{{\rm{vu}}} \mathbf{k}_{\rm{u}} \\
& V_{\hat{\rm{x}} \hat{\rm{x}}_k}= \mathcal{H}_{\hat{\rm{x}} \hat{\rm{x}}} + \mathbf{K}^{\rm{T}}_{\rm{u}} \mathcal{H}_{{\rm{u}}\hat{\rm{x}}} + \mathcal{H}_{\hat{\rm{x}}{\rm{u}}}  \mathbf{K}_{\rm{u}}+ \mathbf{K}^{\rm{T}}_{\rm{v}} \mathcal{H}_{{\rm{v}}\hat{\rm{x}}} + \mathcal{H}_{\hat{\rm{x}}{\rm{v}}}  \mathbf{K}_{\rm{v}} \\
& \ \ \ \ + \mathbf{K}^{\rm{T}}_{\rm{u}} \mathcal{H}_{{\rm{uu}}} \mathbf{K}_{\rm{u}} + \mathbf{K}^{\rm{T}}_{\rm{v}} \mathcal{H}_{{\rm{vv}}} \mathbf{K}_{\rm{v}} + \mathbf{K}^{\rm{T}}_{\rm{u}} \mathcal{H}_{{\rm{uv}}} \mathbf{K}_{\rm{v}} + \mathbf{K}^{\rm{T}}_{\rm{v}} \mathcal{H}_{{\rm{vu}}} \mathbf{K}_{\rm{u}}
\end{split} \end{align*}
with terminal conditions $V_N = \phi(\hat{\rm{x}}_N), \ V_{\hat{\rm{x}}_N} = \phi_{\hat{\rm{x}}}(\hat{\rm{x}}_N)$ and $\ V_{\hat{\rm{x}}\hat{\rm{x}}_N} = \phi_{\hat{\rm{x}}\hat{\rm{x}}}(\hat{\rm{x}}_N)$. Finally, the safety embedded system is propagated forward using the dynamics \eqref{safety embedded game discrete control system} given the optimal polices as ${\rm{u}}_k = \bar{{\rm{u}}}_k + \delta {\rm{u}}^*_k $ and ${\rm{v}}_k = \bar{{\rm{v}}}_k + \delta {\rm{v}}^*_k$. This is repeated until convergence is achieved.

\subsubsection{Regularization}
In the DDP literature, to avoid irregularities while inverting the matrices in the backward pass and to ensure the cost reduction by the min player and accession by the max player, one may add a regularization term to $\mathcal{H}_{\rm{{\rm{uu}}}}, \mathcal{H}_{\rm{ux}}, \mathcal{H}_{\rm{vv}}, \mathcal{H}_{\rm{vx}}$ \cite{liao1991convergence} which was adopted in \cite{Ogunmolu2018minmaxiDG} for their iterative Dynamic Game (iDG). Another option is to use eigen decomposition to ensure that the eigenvalues of each matrix is proper as in \cite{Todorov2005GeniLQG} which was adopted in \cite{sun2018minmax} for their GT-DDP.

\subsubsection{Improved Line-Search for Min-Max Optimal Control}
In optimization, line-search is a classical method used to select a proper step size (also known as learning rate) towards the decent/ascent direction. In the DDP literature, a backtracking search is performed on the feedforward term of the optimal controller through a scaling parameter $\alpha$ \cite{tassa2012synthesis,Tassa2014controllimited} to find a proper local cost reduction. In the min-max DDP literature \cite{morimoto2003minimax,Ogunmolu2018minmaxiDG,sun2018minmax}, no such procedure was reported besides regularization. %It appears that previous works either did not perform proper decent and ascent adaptive step lengths during line-search at all or performed one on the min-player only, as in standard DDP. 
For safety-critical applications, systems' constraints make the optimization problem more complicated and harder to solve. Therefore, for such a highly nonlinear control problem, as advocated in \cite{tassa2012synthesis}, without a proper step toward the decent/ascent direction, a poor and possibly divergent solution is inevitable even with regularization on the controllers matrices. Moreover, with a poor step, irregularities are expected which call for regularization that would not guarantee convergence. Nonetheless, with an adequate step, the algorithm has a higher chance to converge and may not need explicit regularizations.

From a game theoretic view, the gain of one player is a loss for the other. When each player plays its optimal strategy against the other's, the game reaches a saddle-point, when it exists. In \textit{min-max} optimal control settings, the differential game interpretation can take the Stackleberg game, also known as a leader-follower game, since we can assume that the max player maximizes first and then the min player minimizes given that the min player can observe the max player strategy at that time instant \cite{Ball1989h_infandgames}. The min player does not observe future strategies of the max player, however. Therefore, we propose line-searching an adequate cost increase step by the max player, then line-searching an adequate cost decrease step by the min player.

To our interest, preforming a proper line-search for both players helps converging to the optimal solution and hence robustifiying our controller. Indeed, performing line-search for the max player makes it more aggressive in the sense that it tries to find the optimal strategy to increase the cost which then the min player observes and optimizes about hoping to achieve a saddle point with a more robust control. Min-max DDP \cite{Ogunmolu2018minmaxiDG,sun2018minmax} with regularization only, solves the unconstrained problems with no issues. However, once the constraints are imposed, a more complex problem is faced, and the algorithm fails to find a convergent solution.

The performed backtracking search takes the form
\begin{equation}
\vspace{-1mm}
   \delta {\rm{v}}_k^* = \alpha_{v} \mathbf{k}_{{\rm{v}}_k} + \mathbf{K}_{{\rm{v}}_k} \delta \hat{\rm{x}}_k, \ 
   \delta {\rm{u}}_k^* = \alpha_u \mathbf{k}_{{\rm{u}}_k} + \mathbf{K}_{{\rm{u}}_k} \delta \hat{\rm{x}}_k 
\vspace{-0.25mm}
\end{equation}
where $\alpha_v$ and $\alpha_u$ are the search parameters, which start with a value of $1$ and are iteratively reduced as needed. As in \cite{tassa2012synthesis}, we use the expected total cost change 
\begin{equation}\begin{split}
    & \Delta J(\alpha_u, \alpha_v) = \sum_{k=1}^{N-1} \alpha_u \mathbf{k}_{{\rm{u}}_k}^{\rm{T}} \mathcal{H}_{{\rm{u}}_k} + \alpha_v \mathbf{k}_{{\rm{v}}_k}^{\rm{T}} \mathcal{H}_{{\rm{v}}_k} + \\
    & \alpha_v \alpha_u \mathbf{k}_{{\rm{u}}_k}^{\rm{T}} \mathcal{H}_{{\rm{uv}}_k} \mathbf{k}_{{\rm{v}}_k} + \frac{1}{2} \big( \alpha_u^2 \mathbf{k}_{{\rm{u}}_k}^{\rm{T}} \mathcal{H}_{{\rm{uu}}_k} \mathbf{k}_{{\rm{u}}_k} + \alpha_v^2 \mathbf{k}_{{\rm{v}}_k}^{\rm{T}} \mathcal{H}_{{\rm{vv}}_k} \mathbf{k}_{{\rm{v}}_k}  \big)
\end{split} \end{equation}
The solution is accepted when the ratio of the actual change in cost to the expected one, $z = \frac{\Delta V}{\Delta J}$, is positive when minimizing and is negative when maximizing. The proposed algorithm is summarized in Algorithm 1.

%\begin{remark}
%\textcolor{blue}{Note: Added line-search on the feed-forward gains for the maximizer and the minimizer which gave a better performance in terms of convergence (without this, sometimes no convergence or results in positive $Q_{vv}$ which asks for regularization that doesn't always work/help. Line-search is usually done for the min problem, but not done in the literature for min-max. The line-search appears to change the behavior of the maximizer a lot more in the inverted pendulum example so far (it seems to be more aggressive)! I believe this helps the min player $u$ being more robust! This is validated as the one without line-search is less robust and couldn't successfully handle the uncertainty in the examples. On the other hand, the solution with the line-search is more robust to uncertainties!}
%\end{remark}
\begin{algorithm} [ht] 
\SetAlgoLined
 \textbf{Input:} Model $f$, initial condition ${\rm{x}}_0$, running cost $\mathcal{L}$, terminal cost $\phi$, safe region's function $h$, nominal controls $\bar{{\rm{u}}}, \ \bar{{\rm{v}}}$, horizon $N$, convergence threshold $\epsilon$\;
 \textbf{Output:} $V^*, {\rm{x}}^*, \ \mathbf{k}_{\rm{u}}, \ \mathbf{K}_{\rm{u}}, \ {\rm{u}}^*, \ \mathbf{k}_{\rm{v}}, \mathbf{K}_{\rm{v}}, \ {\rm{v}}^*$\;
 \textbf{Precompute:} Barrier state dynamics $f_w$ and create $\hat{f}$, nominal trajectory $\bar{\hat{\rm{x}}}$\;
 \While{$\Delta V > \epsilon$}{
 Compute costs $\mathcal{L}, \ \phi$ % given $(\bar{\hat{\rm{x}}}, \bar{\rm{u}}, \bar{\rm{v}})$
 and $V, \ V_{\hat{x}}, \ V_{\hat{x}\hat{x}}$ at $k=N$\;
 \For{$k=N-1$ to $1$}{
 Compute $\hat{f}_{\hat{\rm{x}}},\ \hat{f}_{{\rm{u}}},\ \hat{f}_{{\rm{v}}}$\;
 Compute the matrices in \eqref{H matrices}\;
 Regularize $\mathcal{H}_{{\rm{uu}}}$ and $\mathcal{H}_{{\rm{vv}}}$ if needed\;
 Compute $\mathbf{k}_{\rm{u}}, \ \mathbf{K}_{\rm{u}}, \ \mathbf{k}_{\rm{v}},\ \mathbf{K}_{\rm{v}}, \ V_{\hat{x}}, \ V_{\hat{x}\hat{x}}$\;
 }
 \For{$k=1$ to N-1}{
 Compute $\delta {\rm{v}}^*$ and ${\rm{v}}^*= \bar{{\rm{v}}} + \delta {\rm{v}}^*$\;
  Backtracking line-search $\alpha_v$ and update $\rm{v}^*$ given $(\bar{\hat{\rm{x}}},\bar{\rm{u}}, {\rm{v}}^*) $\;
 Compute $\delta {\rm{u}}^*$ and ${\rm{u}}^*= \bar{{\rm{u}}} + \delta {\rm{u}}^*$\;
 Backtracking line-search $\alpha_u$ and update $\rm{u}^*$ given $(\bar{\hat{\rm{x}}},{\rm{u}}^*, {\rm{v}}^*)$\;
 Forward propagate $\hat{f}(\bar{\hat{\rm{x}}},{\rm{u}}^*, {\rm{v}}^*) $\; 
 }
 Update $\Delta V, \ \bar{\hat{{\rm{x}}}}, \ \bar{{\rm{u}}}, \ \bar{{\rm{v}}}$
 }
\caption{Safety Embedded Min-Max DDP} \label{Safety Embedded Min-Max DDP algorithm}
\end{algorithm}

\begin{table*}[ht]
\vspace{-2mm}
\caption{Comparison of MinMax DBaS-DDP (proposed) and DBaS-DDP (basline) under different uncertainty or disturbance levels. Safety, reachability and success rates are the percentages of trajectories that satisfy the safety constraints, reach the target, and safely reach the target respectively. RMSD is the root-mean-square deviation from reaching the target and total state variance is the sum of variances of all states over the entire time horizon.}
\vspace{-2mm}
\centering
\begin{tabular}{|c|cccc|cccc|}
\hline
System & \multicolumn{4}{c|}{Pendulum}& \multicolumn{4}{c|}{Quadrotor}     \\ \hline
Noise Level & \multicolumn{2}{c|}{Moderate}& \multicolumn{2}{c|}{High}   & \multicolumn{2}{c|}{Moderate}& \multicolumn{2}{c|}{High} \\ \hline
Algorithm & \multicolumn{1}{c|}{Proposed} & \multicolumn{1}{c|}{Baseline}   & \multicolumn{1}{c|}{Proposed}& \multicolumn{1}{c|}{Baseline} & \multicolumn{1}{c|}{Proposed} & \multicolumn{1}{c|}{Baseline}   & \multicolumn{1}{c|}{Proposed}& \multicolumn{1}{c|}{Baseline}
\\ \hline
Safety ($\%$)   & \multicolumn{1}{c|}{\textbf{98.8}} & \multicolumn{1}{c|}{79.4}  & \multicolumn{1}{c|}{\textbf{81.7}} & 59.5  & \multicolumn{1}{c|}{\textbf{98.6}} & \multicolumn{1}{c|}{90.8}  & \multicolumn{1}{c|}{\textbf{94.3}} & 83.3 \\
Reachability ($\%$)   & \multicolumn{1}{c|}{86.7} & \multicolumn{1}{c|}{\textbf{95.6}}  & \multicolumn{1}{c|}{71} & \textbf{79.7}  & \multicolumn{1}{c|}{98.1} & \multicolumn{1}{c|}{\textbf{99.4}}  & \multicolumn{1}{c|}{86.5} & \textbf{90.4} \\
Success ($\%$)  & \multicolumn{1}{c|}{\textbf{86.4}} & \multicolumn{1}{c|}{75.5}  & \multicolumn{1}{c|}{\textbf{66.3}} & 48.9  & \multicolumn{1}{c|}{\textbf{96.7}} & \multicolumn{1}{c|}{90.3}  & \multicolumn{1}{c|}{\textbf{81.6}} & 75.2 \\
RMSD            & \multicolumn{1}{c|}{0.202}    &   \multicolumn{1}{c|}{\textbf{0.119}}  & \multicolumn{1}{c|}{0.275} & \textbf{0.209}  & \multicolumn{1}{c|}{0.655} & \multicolumn{1}{c|}{\textbf{0.584}}  & \multicolumn{1}{c|}{0.852} & \textbf{0.791} \\
Total State Variance & \multicolumn{1}{c|}{\textbf{39.8}}    &   \multicolumn{1}{c|}{86.3}         & \multicolumn{1}{c|}{\textbf{172.9}} & 209.9  & \multicolumn{1}{c|}{\textbf{230.2}}             & \multicolumn{1}{c|}{320.3}  & \multicolumn{1}{c|}{\textbf{421.8}} & 603.5 \\
%Total State \& Input Variance       & \multicolumn{1}{c|}{\textbf{582.5}}   &   \multicolumn{1}{c|}{763.5} & \multicolumn{1}{c|}{\textbf{1313}} & 1706.8 & \multicolumn{1}{c|}{??}             & \multicolumn{1}{c|}{??}  & \multicolumn{1}{c|}{??} & ?? \\
\hline
\end{tabular}
\vspace{-5mm}
\label{tab: full statistical comparison of MinMax-DBaS-DDP and DBaS-DDP}
\end{table*}

\section{Application Examples} \label{Section: Simulation Examples}
In this section, we implement the proposed algorithm, MinMax DBaS-DDP, and compare it against DBaS-DDP \cite{almubarak2021safeddp}. The two methods are compared in terms of safety (defined as not violating the safety constraints), reachability (defined as reaching a terminal state within a pre-specified distance from the target), success (defined as safely reaching the target), root-mean-square deviation from reaching the exact target state and total state variance (defined as the sum of variances of all states over the entire time horizon). We consider two scenarios for each problem, one with a moderate disturbance level and one with a high disturbance level. The numerical results are provided in \autoref{tab: full statistical comparison of MinMax-DBaS-DDP and DBaS-DDP}. To get meaningful results, each experiment is run for $1000$ Monte Carlo simulations. In both examples, and for both cases of noise levels, MinMax DBaS-DDP consistently achieves the highest safety and success rates with lower variance but that comes at the expense of having a bigger RMSD from the target, which means slightly less reachability rate. It is worth mentioning that min-max DDP \textit{with} barrier states but without the proposed line-search as the algorithms in \cite{Ogunmolu2018minmaxiDG,sun2018minmax} fails to converge or compute meaningful solutions as we hypothesised. Moreover, it is worth mentioning that converging to a meaningful saddle point, i.e. a solution that provides a robust feedback policy, using an iterative local algorithm for such a constrained differential game may not be easy. Hence, a dense line-search could be needed. 

All systems are initialized with steady-state nominal trajectories (zero input for the inverted pendulum and hovering over the initial condition for the quadrotor). Both systems are discretized using the forward Euler method with $\Delta t = 0.01$.

%\begin{table}[ht]
%\centering
%\begin{tabular}{|c|cccc|cccc|}
%\hline
%System & \multicolumn{2}{c|}{Pendulum}& \multicolumn{2}{c|}{Quadrotor}     \\ \hline
%Noise Level & \multicolumn{1}{c|}{Moderate}& \multicolumn{1}{c|}{High}   & \multicolumn{1}{c|}{Moderate}& \multicolumn{1}{c|}{High} \\ \hline
%Safety   & \multicolumn{1}{c|}{19.4} & \multicolumn{1}{c|}{26.6}  & \multicolumn{1}{c|}{7.8} & 11   \\
%Reachability   & \multicolumn{1}{c|}{-8.9} & \multicolumn{1}{c|}{-8.7}  & \multicolumn{1}{c|}{-1.3} & -3.9   \\
%Success & \multicolumn{1}{c|}{10.9} & \multicolumn{1}{c|}{18.4}  & \multicolumn{1}{c|}{6.4} & 6.4 \\
%RMSD      & \multicolumn{1}{c|}{-69.9} & \multicolumn{1}{c|}{-26.7}  & \multicolumn{1}{c|}{-12.2} & -7.6 \\
%Total State Variance & \multicolumn{1}{c|}{53.9} & \multicolumn{1}{c|}{17.6}  & \multicolumn{1}{c|}{28.1} & 30.1 \\
%%Total State \& Input Variance Reduction ($\%$)  & \multicolumn{1}{c|}{23.7} & \multicolumn{1}{c|}{23.1}  & \multicolumn{1}{c|}{??} & ?? \\
%\hline
%\end{tabular}
%\vspace{1mm}
%\caption{Improvements and Reductions Percentages Summary of \autoref{tab: full statistical comparison of MinMax-DBaS-DDP and DBaS-DDP}.}
%\label{tab: summary of improvements of the comparison of MinMax-DBaS-DDP and DBaS-DDP}
%\end{table}

\subsection{Velocity Constrained Inverted Pendulum}
We start with a constrained simple inverted pendulum to get a good comprehension about the effect of considering a maximizing player to the safety-critical problem. Consider the inverted pendulum dynamics
\begin{equation} \label{inverted pendulum dynamics} \vspace{-0.5mm}
    I \ddot{\theta} + b \dot{\theta} - m g l \sin(\theta) = {\rm{u}} + {\rm{v}}
\vspace{-0.5mm}
\end{equation}
with a safety constraint in the angular velocity not to exceed $5$, i.e. $|\dot{\theta}| < 5 \ \text{rad}/\text{sec}$. We assume that the model that we use to design our controller has $l=0.75$ m, $b=0.15$ N $\cdot$ s/m, $m = 1.5$ kg, $I = m l^2$ and $g = 9.81$ m/s$^2$. The model's parameters are assumed to be off from the true system. We consider two scenarios in which for the first scenario, the uncertainty is moderate which is modeled by a normal distribution with mean $\mu_{\text{Moderate}}=10\%$ and standard deviation $\sigma_{\text{Moderate}}=30\%$. For the second scenario, the uncertainty is high which is modeled by a normal distribution with mean $\mu_{\text{High}}=20\%$ and standard deviation $\sigma_{\text{High}}=50\%$. Moreover, we assume that each model parameter has a different uncertainty, i.e. $l_{\text{true}} = 0.75 c_1$m, $b_{\text{true}}=0.15c_2$ N $\cdot$ s/m and $m_{\text{true}}=1.5c_3$ kg, where $c_i=1-\mathtt{x}$, where $\mathtt{x} \sim \mathcal{N} (\mu, \sigma)$ and $i=1,2,3$. The goal is to swing up the pendulum to the up right position $\theta=0$ in one and a half seconds and finish the task with a small angular velocity. The target is \textit{reached} if the final angle is within $0.3$ rad from the target $0$. Forming the problem as an optimal control problem while considering a hostile disturbance, we consider the quadratic cost function
\begin{equation*} 
%{\rm{x}}_k^{\rm{T}} Q {\rm{x}_k} 
     J =  \min_{\rm{u}} \max_{\rm{v}} \sum_{k=1}^{N-1} \big(R_{\rm{u}} {\rm{u}}_k^2 - R_{\rm{v}} {\rm{v}}_k^2 \big) + {\rm{x}}_N^{\rm{T}} S {{\rm{x}}_N} 
\end{equation*}
subject to \eqref{inverted pendulum dynamics} and $|\dot{\theta}| < 5 \ \text{rad}/\text{sec}$. To address the safety constraint, we define $h =5^2 - \dot{\theta}^2$ and create the barrier function $\beta = \frac{1}{5^2 - \dot{\theta}^2}$ and finally we define a DBaS, $\rm{w}$, according to \eqref{dbas dynamics}. Embedding the DBaS into the dynamics gives a third order system with the cost function
\begin{equation} \label{inv pend quad cost}
     J =  \min_{\rm{u}} \max_{\rm{v}} \sum_{k=1}^{N-1} \big( Q_{\text{DBaS}} {\rm{w}}_k^2 + R_{\rm{u}} {\rm{u}}_k^2 - R_{\rm{v}} {\rm{v}}_k^2 \big) + {\rm{\hat{x}}}_N^{\rm{T}} S {{\rm{\hat{x}}}_N} 
\end{equation}
where we choose $Q_{\text{DBaS}} = 1000, R_{\rm{u}} = 0.1, R_{\rm{v}} = 1.1, S=\text{diag}(1000,5,500)$. In such a problem, it must be noted that the barrier state uses the poor model and is propagated based on the poor model. 

\begin{figure} [htb]
    \centering
    \vspace{-3mm}
    \subfloat{\includegraphics[trim=10 0 10 0, clip, width=.84\linewidth]{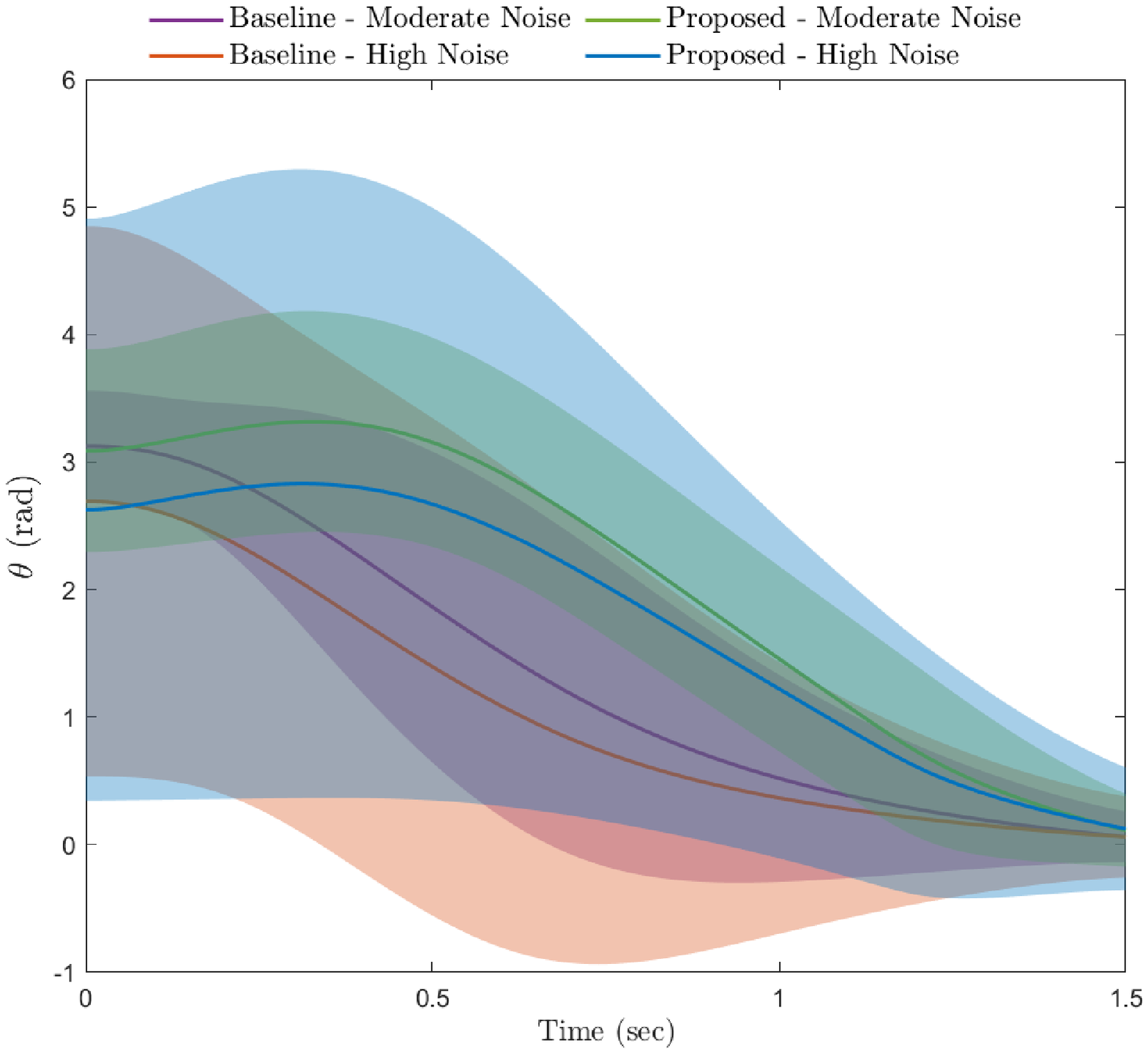}}
    \\
    \vspace{-6mm}
    \subfloat{\includegraphics[trim=10 0 10 20, clip, width=.84\linewidth]{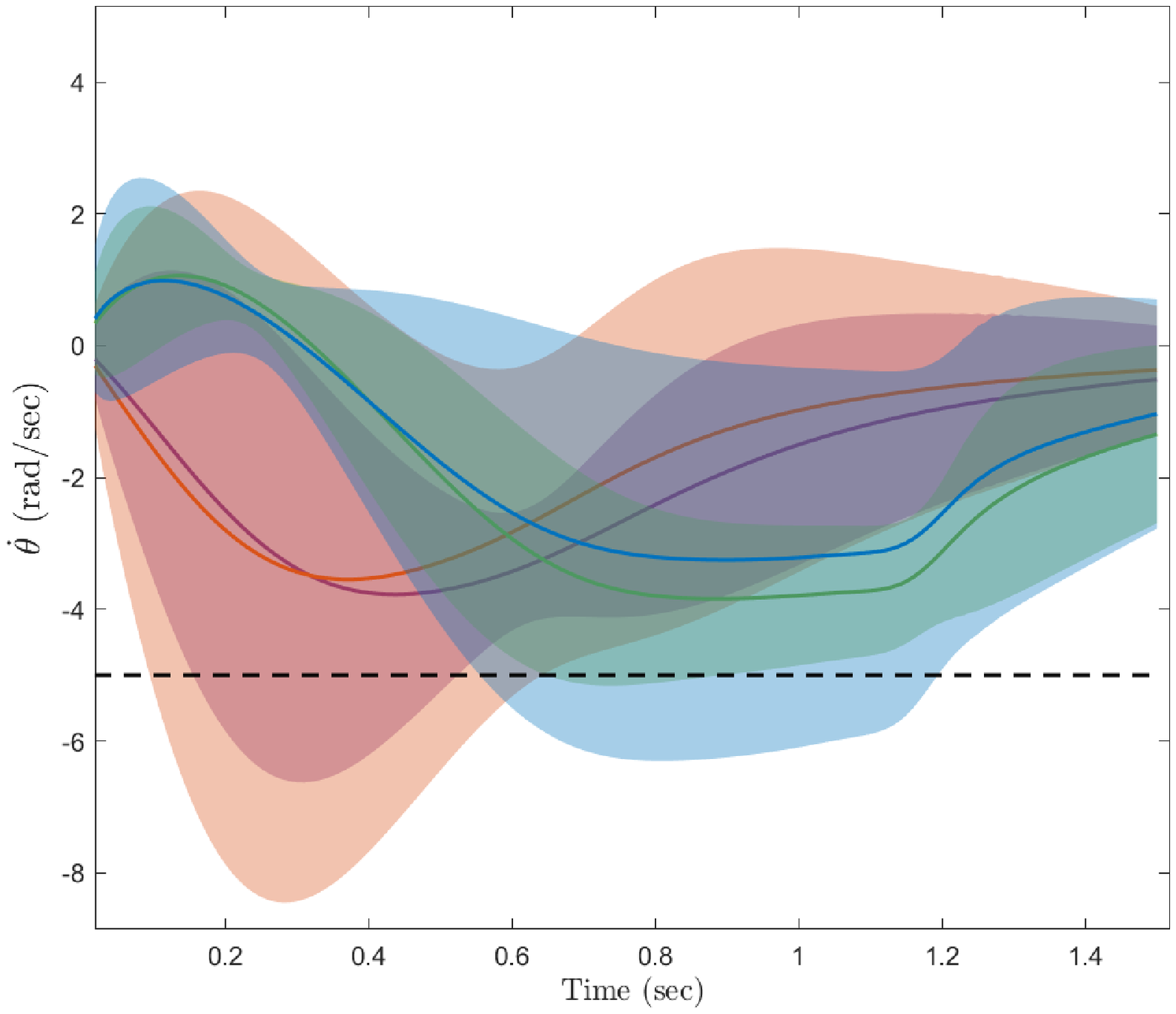}}
    \\
    \vspace{-6mm}
    \subfloat{\includegraphics[trim=10 0 10 20, clip, width=.84\linewidth]{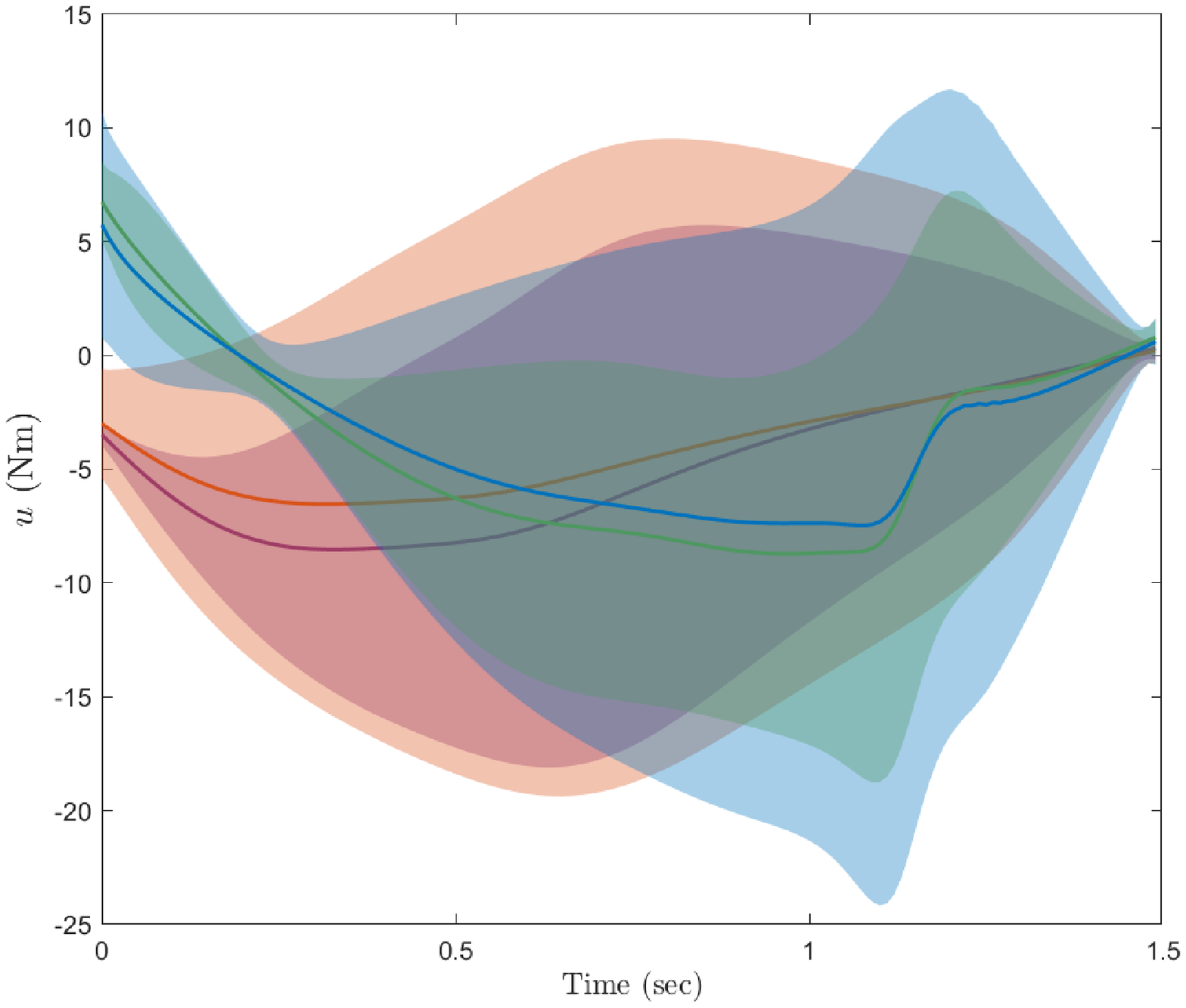}}
    \vspace{-5mm}
    \caption{Inverted pendulum states and control under DBaS-DDP (basline) and MinMax-DBaS-DDP (proposed) with moderate and high parametric uncertainties in the model. Solid trajectories represent the mean trajectories and the shaded regions represent confidence regions of $95\%$. It can be seen that the proposed approach provides a more robust and safer feedback control, i.e. less variance and less violation of the safety bound (black dashed line) compared to the standard (min) approach.}
    \label{fig:minmaxddp_inv_pend}
    \vspace{-8mm}
\end{figure}

The results are As shown in \autoref{fig:minmaxddp_inv_pend} and in \autoref{tab: full statistical comparison of MinMax-DBaS-DDP and DBaS-DDP}. It can be concluded that the proposed MinMax-DBaS-DDP clearly robustifies the controller in terms of handling uncertainty and completing the task safely. Nonetheless, this comes at the sacrifice of having a larger RMSD, which can be seen in \autoref{fig:minmaxddp_inv_pend} near the end of the trajectory that it has a larger variance. 

\subsection{Quadrotor in a Windy Constrained Environment}
We consider a quadrotor flying in a windy environment with obstacles. We use the model derived in \cite{Sabatino2015QuadrotorCM} in which the force of a sinusoidal wind turbulence enters the linear velocities, in the body frame, of the quadrotor in all three dimensions $x,y$ and $z$ randomly with a zero mean and a standard deviation $\sigma$, i.e. $F_{i-\text{wind}}=\sigma \rho \sin(t)$ where $F$ is the force, $i$ is $x,y$ or $z$ and $\rho$ is a random variable drawn from the standard normal distribution. The obstacle course consists of randomly generated obstacles and thus the safe set is defined as $S^{\circ} =\{ [x \ y \ z]^{\rm{T}} \in \mathbb{R}^3 | \ (x - o_{j_x})^2 + (y - o_{j_y})^2 + (z - o_{j_z})^2 - r_j^2 > 0\}$ where $j$ is number of obstacles, and $o_j$ and $r_j$ are the $j$ obstacle's center and radius. The quadrotor is to fly from the starting point $(10,0,-1)$ to the target point $(-5,-3,2)$ in a moderate turbulence with $\sigma=15$ and a high turbulence with $\sigma=20$ in $5$ seconds. The target is considered \textit{reached} if the final position is within $2$ units from it.

We first examine different robustness results with different choices of $R_{\rm{v}}$ under the moderate turbulence. For a quadratic cost as in \eqref{inv pend quad cost} and one DBaS representing all the obstacles, we choose the parameters, $R_{\rm{u}}=10^{-4}I, Q_{\text{DBaS}} = 0.1, S=I, S_{i}=10$. \autoref{Fig:Quadrotor with two different Rv under wind sin distrubances} shows two robustness results for two different penalization coefficients of the maximizing player's input, $R_{\rm{v}}=\frac{1}{100}I$ (blue) and $R_{\rm{v}}=\frac{1}{150}I$ (purple). The solid trajectories represent the mean trajectories and the shaded regions represent confidence regions of $95\%$ generated by $1000$ trajectories. It can be seen that the control policy generated by the min player with less penalization of the max player generates a more risk sensitive solution and is completely safe.

\begin{figure} [htb]
    \hspace*{-6mm}
    \subfloat{\includegraphics[trim=60 5 20 20, clip, width=.7\linewidth]{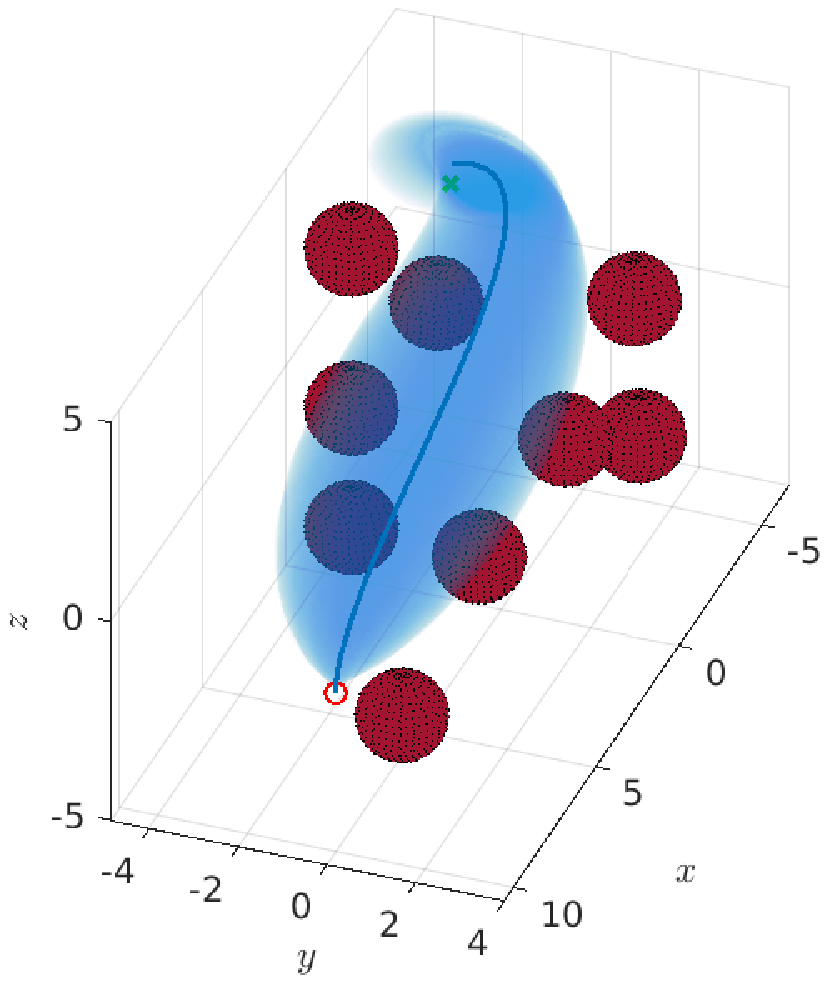}}
    \hspace*{-1.5cm}
    \subfloat{\includegraphics[trim=60 5 20 20, clip, width=.7\linewidth]{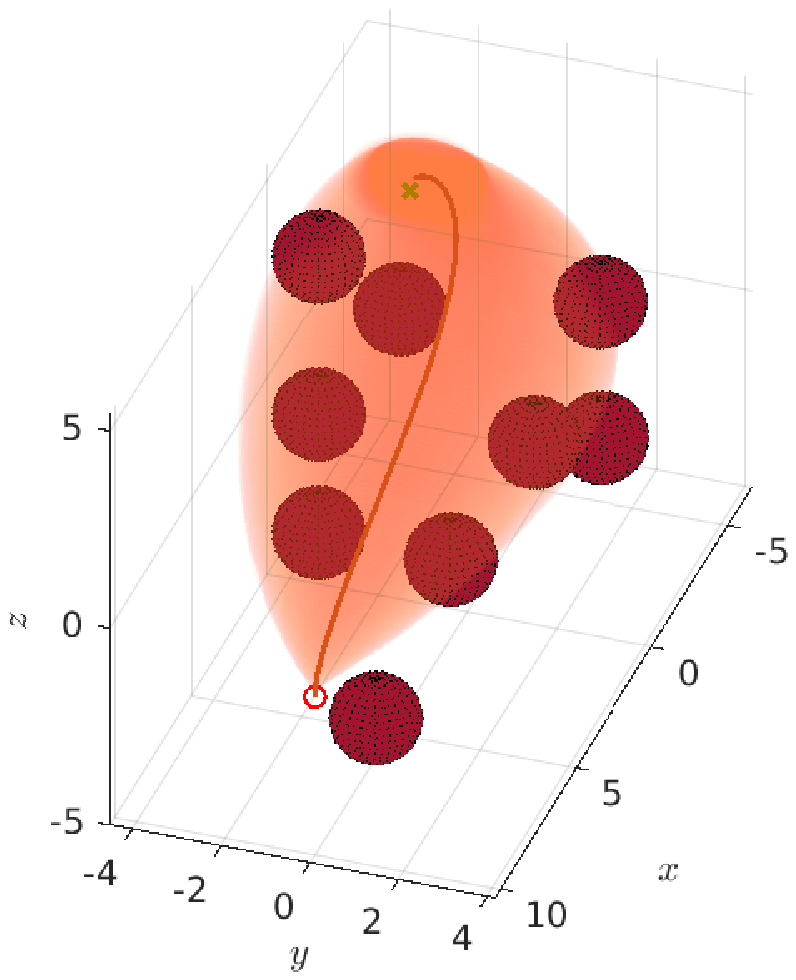}}
    \\
    \hspace*{-6mm}
    \subfloat{\includegraphics[trim=60 5 20 20, clip, width=.7\linewidth]{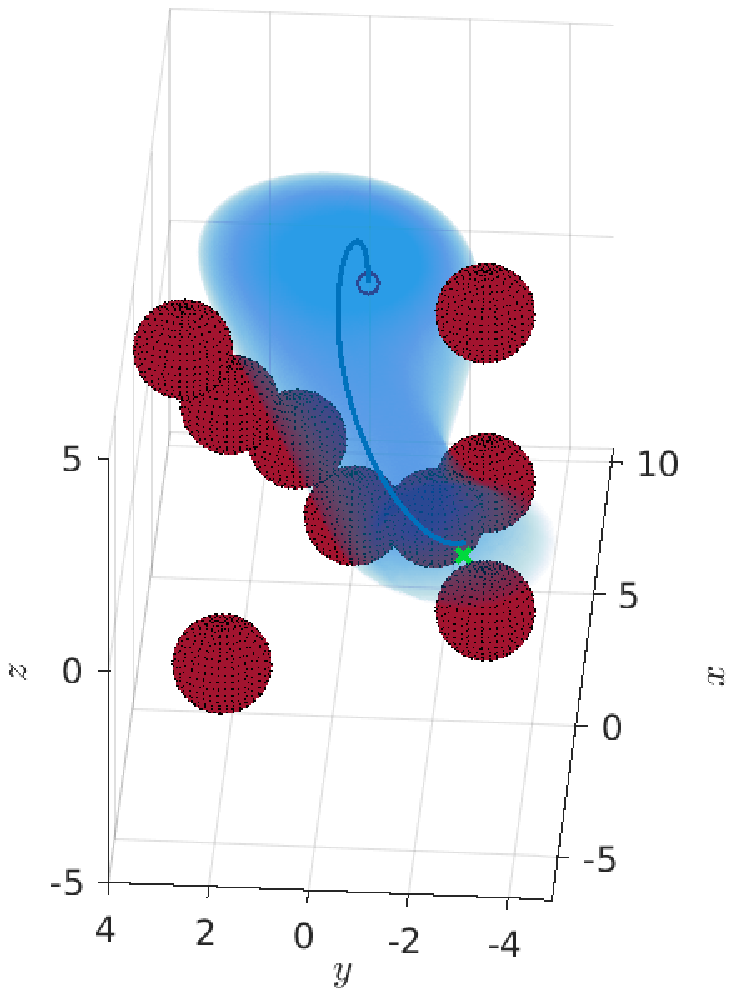}}
    \hspace*{-1.5cm}
    \subfloat{\includegraphics[trim=60 5 20 20, clip, width=.7\linewidth]{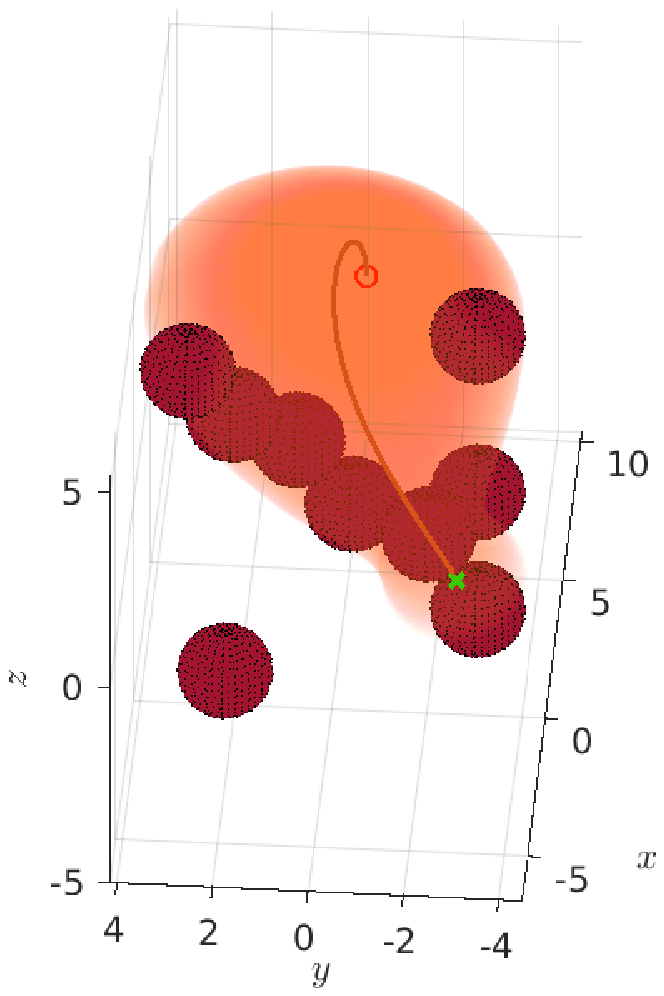}}
     \\
    \hspace*{-6mm}
    \subfloat{\includegraphics[trim=60 5 20 20, clip, width=.675\linewidth]{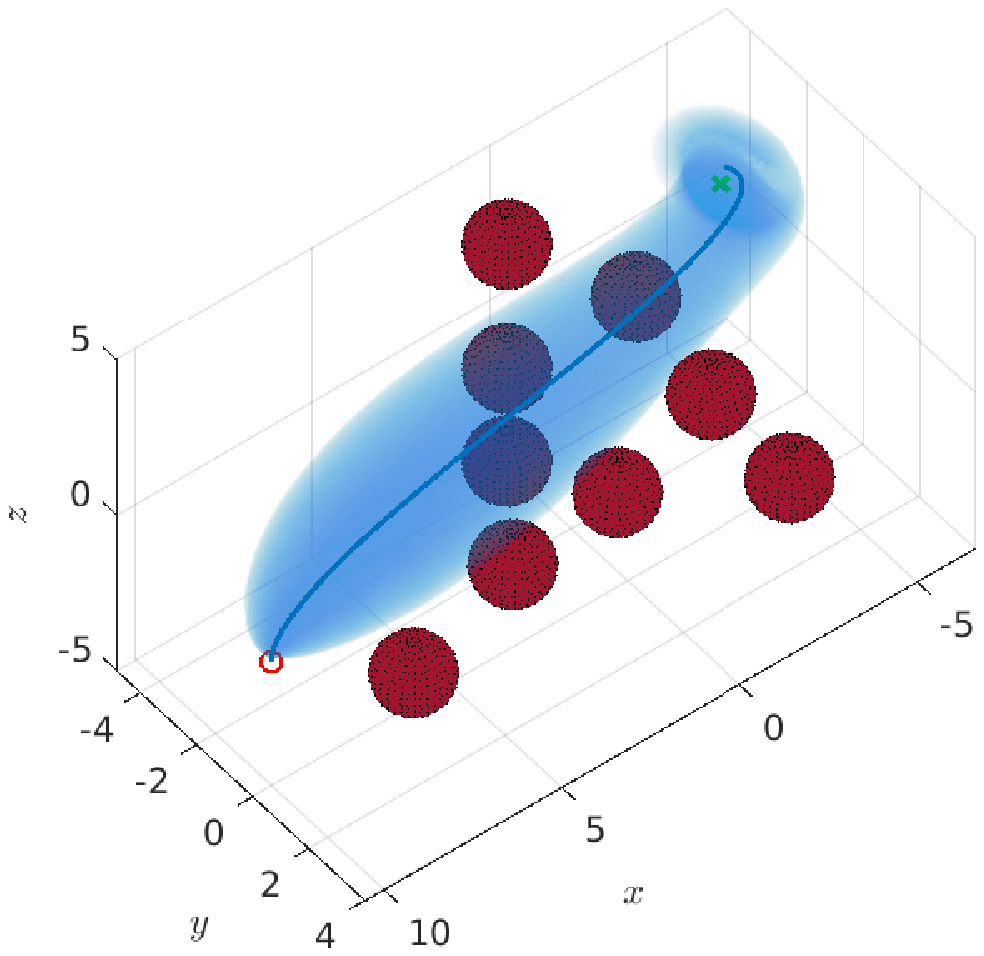}}
    \hspace*{-1.1cm}
    \subfloat{\includegraphics[trim=60 5 20 20, clip, width=.675\linewidth]{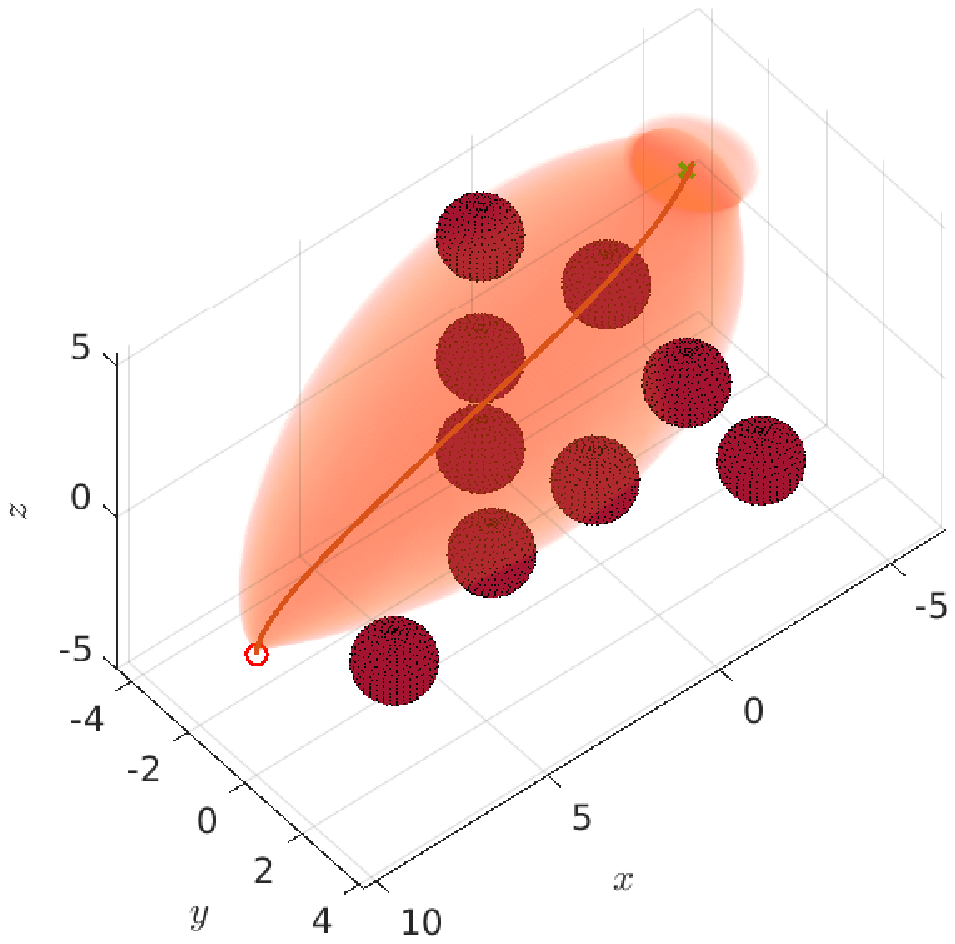}}
    \caption{Three angles view of the quadrotor flying in a windy environment using MinMax-DBaS-DDP (left, in blue) and DBaS-DDP (right, in orange). The figures show the mean trajectories (solid) and the $90\%$ confidence region (shaded). Clearly, the proposed controller is able to \textit{safely} fly the quadrotor to avoid the obstacles (dark red ellipsoids) and reach the target (green \rm{x}) with less variance. %The bottom figures show $500$ trajectories safely reaching the target point and the corresponding confidence interval of the DBaS, which is bounded implying safe solutions.
    }
    \label{fig:minmaxddp_quad90cr_comparison}
    \vspace{-5mm}
\end{figure}

Now we compare the proposed approach against the standard DBaS-DDP while allowing less control authority for the minimizing player. Namely, we pick $R_{\rm{u}}=10^{-2}I$ and $R_{\rm{v}}=15 \times 10^{-2}$, which is the smallest coefficient obtained with a convergent solution. \autoref{fig:minmaxddp_quad90cr_comparison} shows a comparison between the proposed MinMax-DBaS-DDP and the DBaS-DDP flying the quadrotor in the moderately turbulent environment. As shown, the proposed controller provides a substantial improvement in robustness compared to the standard DBaS-DDP. The standard DBaS-DDP has a larger, hazardous confidence region while the more robust MinMAx-DBaS-DDP, has a tighter and safer confidence region. \autoref{tab: full statistical comparison of MinMax-DBaS-DDP and DBaS-DDP} provides the detailed comparison under the moderate disturbance as well as the high disturbance in which it is shown that the proposed method achieves higher safety and success rates and less variance but a lower reachability rate and a larger RMSD.

% \autoref{tab: full statistical comparison of MinMax-DBaS-DDP and DBaS-DDP} provides the details for the experiments in the moderately and highly turbulent environments. We also tested the two algorithms with fixed wind forces and as expected, the standard DBaS-DDP can handle small forces but not moderate to large ones unlike the proposed MinMax-DBaS-DDP.

From our experiments, we can conclude that penalizing $R_u$ less or $R_v$ more tends to ease the problem for the min player and thus converges quickly, but makes the controller less robust. However, to increase robustness, the max player needs to be penalized less but needs to be tuned carefully to avoid divergence especially with the existence of GT-DBaS as we mentioned earlier. Moreover, penalizing the running and terminal states more, including the GT-DBaS, tends to decrease robustness of the min player. This can be interpreted as increasing the importance of achieving the target over handling disturbances. In other words, there is a trade-off between increasing robustness and completing the task with a small deviation from the target. 

\vspace{-1mm}
\section{Conclusion} \label{Section: Conclusion}
In this paper, a robust and safe trajectory optimization algorithm was presented. To enforce safety, the proposed algorithm utilized barrier states that are embedded into the model of the safety-critical system to reconstruct the constraints as performance objectives in a higher dimensional state space. For robustness, a min-max optimal control approach was adopted utilizing a game-theoretic interpretation to the problem developing a more robust minimizing controller. The min-max optimal control problem was solved using differential dynamic programming. Finally, an improved line-search strategy in the feed-forward gains of players was proposed. The line-search helps the max player to increase the cost and helps avoiding irregularities in $\mathcal{H}_{{\rm{vv}}}$ and $\mathcal{H}_{{\rm{uu}}}$. This results in a more risk-aware min player, and hence a robust controller. Two application examples representing parametric and non-parametric uncertainties were presented. The algorithm was compared against standard DBaS-DDP in which the proposed method was shown to consistently provides more robust and safer control but with a larger RMSD from the target, which implies a smaller reachability rate.  

Future work will include developing a receding-horizon min-max DDP to be deployed on physical systems and extending the work to decentralized multi-agent control and planning. Generalizing the algorithm to risk-sensitive stochastic trajectory optimization is an active research.

\section{Acknowledgment}
The authors would like to thank the members of the Autonomous Control and Decision Systems (ACDS) Lab at Georgia Tech especially Yuichiro Aoyama and Bogdan Vlahov for the valuable discussions.

\printbibliography

\end{document}